# Design and performance of the front-end readout electronics for LACT

**Changge Zi,**[a,b,*] **Jinfan Chang,**[a] **Fengfan Yang,**[a] **Ming Wei**[c] **and Wei Wei**[a]

[a] *Institute of High Energy Physics, Chinese Academy of Sciences, Beijing 100049, China*

[b] *China Center of Advanced Science and Technology, Beijing 100190, China*

[c] *College of Physics, Sichuan University, Chengdu 610065, China*

*E-mail*: zicg@ihep.ac.cn

ABSTRACT: The Large High Altitude Air Shower Observatory (LHAASO) is sensitive to Very-High-Energy gamma rays. However, its angular resolution limits its ability to resolve the fine spatial structure of some gamma-ray source regions. The Institute of High Energy Physics is constructing the Large Array of Imaging Atmospheric Cherenkov Telescope (LACT). With a target angular resolution of 0.05°, LACT will complement LHAASO's gamma-ray detection capability to study detailed source morphology and the origin and acceleration mechanisms of cosmic rays. The array is designed to comprise 32 Cherenkov telescopes with 6-meter apertures and is located in the LHAASO array at an altitude of approximately 4410 meters. Each telescope has a camera comprising 1616 silicon photomultipliers (SiPMs). The SiPM camera detects Cherenkov light produced by atmospheric showers initiated by Very-High-Energy gamma rays. The readout electronics integrated inside the camera needs to adjust each SiPM's operating voltage according to its breakdown voltage and operating temperature to stabilize its gain. It also needs to sample the SiPM output pulse at high speed to measure the light intensity. To meet the requirements of operation at high altitude and over a wide temperature range, we designed and developed a front-end readout electronics (FEE) board that integrates 16 channels of SiPM gain compensation and high-speed waveform sampling. The gain compensation circuit independently adjusts the operating voltage of each SiPM according to its temperature and monitors the actual operating voltage. Previous tests of the adopted compensation scheme showed SiPM gain drift below 2% from −20 °C to 30 °C. The waveform sampling circuit uses two 14-bit AC9628 analog-to-digital converters, each operating in eight-channel mode. The SiPM pulse waveform is digitized at 250 million samples per second. The integral nonlinearity of the DC response is 0.0709% over a differential DC voltage range of −950 mV to +950 mV at the AC9628 input. Equivalent electrical pulses with a full width at half maximum of 37.42 ns are used to simulate the SiPM output signals. At a differential pulse amplitude of approximately 237 mV at the AC9628 input, the pulse integral area resolution is 0.2725%. This paper presents the circuit design and performance tests of the FEE for the LACT camera.

---

* Corresponding author.

## Contents

## 1. Introduction

The spatial structures and energy spectra of very-high-energy gamma-ray sources provide important information on the origin and acceleration mechanisms of cosmic rays. The Large High Altitude Air Shower Observatory (LHAASO) has discovered multiple very-high-energy gamma-ray sources. Some of these sources have extended structures, and observations with higher angular resolution are needed to resolve their fine spatial structures [1]. The Large Array of Imaging Atmospheric Cherenkov Telescope (LACT) is planned to comprise 32 Cherenkov telescopes with 6-meter apertures. It is being deployed within the LHAASO array at an altitude of approximately 4410 meters, with a target angular resolution of 0.05° [1, 2]. LACT's high-angular-resolution imaging will complement LHAASO's energy spectrum measurements, providing observational information on the detailed structures and radiation mechanisms of very-high-energy gamma-ray sources.

The layout and structure of the LACT telescopes are shown in figure 1. Each telescope mainly consists of a rotation system, an optical system, a silicon photomultiplier (SiPM) camera and a readout electronics system [3]. The rotation system drives the telescope to track the target gamma-ray source. The optical system focuses Cherenkov light onto the focal plane, where the SiPM array converts the light into electrical pulses. The readout electronics amplifies and samples these pulses at high speed. To reduce the attenuation of weak SiPM pulses during long-distance transmission, the electronics is integrated inside the camera. Together with the SiPM array, it forms a high-density, multichannel detection system.

The 1616 SiPM pixels of the LACT camera are divided into 101 sub-clusters [4, 5]. Each sub-cluster mainly comprises 16 SiPM pixels, a preamplifier board (PRB) and a 16-channel front-end electronics (FEE) board. The PRB supports the SiPMs and provides signal shaping.

The FEE independently adjusts the operating voltages of the 16 SiPMs, amplifies and samples their pulse signals at high speed, and monitors their operating temperatures and voltages.

As shown in figure 1, the PRB is located at the front of the camera. Replacing or servicing it requires removal of the optical filter at the front of the camera. In contrast, the FEE can be removed independently from the rear of the camera along its support rails. To simplify the PRB and reduce its maintenance requirements, the signal amplification circuit was moved to the more accessible FEE after the PRB shaping schemes had been tested and a scheme selected. The selected shaping scheme is similar to the adder circuit described in [4]. Its passive resistor-capacitor network is located on the PRB, whereas the amplification stage, implemented with OPA892 amplifiers, is located on the FEE. The PRB also carries 16 analog temperature sensors, one behind each SiPM. Their analog outputs are read by the temperature monitoring analog-to-digital converter (ADC) on the FEE to obtain the operating temperature of each SiPM.

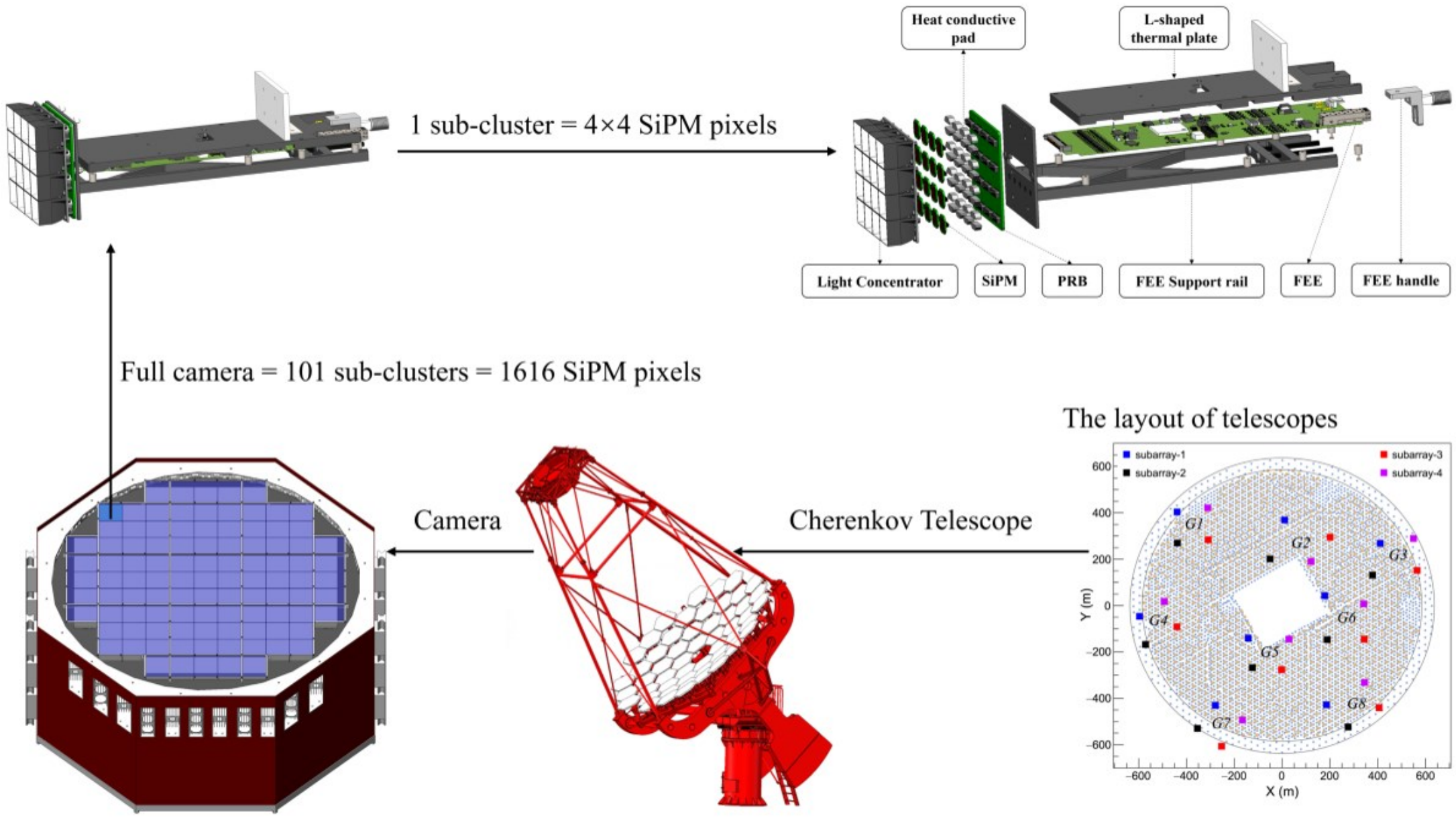

**Figure 1.** Layout of the LACT telescope array and structure of a sub-cluster [2, 4].

The SiPMs on the PRB are located at the front of the camera and are separated from the external environment only by the optical filter. Their operating temperatures are strongly affected by outdoor temperature variations and approximately follow the ambient temperature, ranging from −20 °C to 30 °C. The SiPM breakdown voltage changes with temperature. At a fixed operating voltage, this changes the overvoltage, defined as the operating voltage minus the breakdown voltage, and hence causes gain drift. This drift affects the accuracy of Cherenkov light intensity measurements [6]. In addition, differences in the breakdown voltages of individual SiPMs lead to gain nonuniformity when the same operating voltage is applied. The FEE sets each operating voltage independently using the manufacturer-supplied breakdown voltage parameters and the measured SiPM temperature. This reduces channel-to-channel gain nonuniformity and compensates for the gain temperature drift. The FEE also monitors the actual SiPM operating voltages with an ADC to assess the operating status of the SiPMs.

In the SiPM output waveform, pulses produced by Cherenkov light are superimposed on the baseline produced by night sky background light. The baseline level and its fluctuations can be used to assess the night sky background noise. This information provides a basis for dynamically adjusting the trigger threshold of each channel to reduce false triggers. The high-speed waveform sampling circuit uses DC coupling to record both the pulse and the baseline before and after it [4]. The baseline is determined from intervals without Cherenkov pulses. After baseline subtraction, the pulse is integrated over a selected window, and the integral area is used to represent the Cherenkov light intensity. High-speed waveform sampling also allows the integration window to be optimized to cover the pulse while reducing the contribution of background light to the signal measurement.

The LACT camera moves with the telescope to track target gamma-ray sources. Its mechanical design directly affects the load on the support and rotation systems. At the same time, the multichannel electronics inside the camera continuously generates heat and requires an effective cooling system for stable operation. The thermal design should consider heat removal together with the weight, mechanical complexity and maintenance cost of the cooling system. For example, the prototype Schwarzschild-Couder Telescope, developed for the Cherenkov Telescope Array, uses a liquid cooling system. A chiller cools the coolant, which is pumped into the camera through a feed valve and circulated through the cooling pipes. Fans direct air across the chilled pipes and through the camera electronics [7]. Although this system provides effective cooling, it requires chiller lines, a rotary union and a pressure regulator to accommodate telescope tracking. These components increase the complexity of the mechanical interfaces and the operating and maintenance costs.

LACT combines thermally conductive plates with separate air tubes, which helps simplify the camera's mechanical interfaces and reduce operating costs. As shown in figure 2, air tubes are arranged between the sub-clusters inside the camera. Heat generated by the electronics is conducted through aluminum L-shaped thermal plates to the walls of the air tubes. Fans drive ambient air through the tubes to remove the heat [5]. The tube walls separate the external cooling airflow from the electronics compartment, preventing external air from entering the compartment directly. This arrangement allows heat removal while keeping the camera sealed. The sealed structure effectively reduces the ingress of dust and moisture that could damage the electronic components.

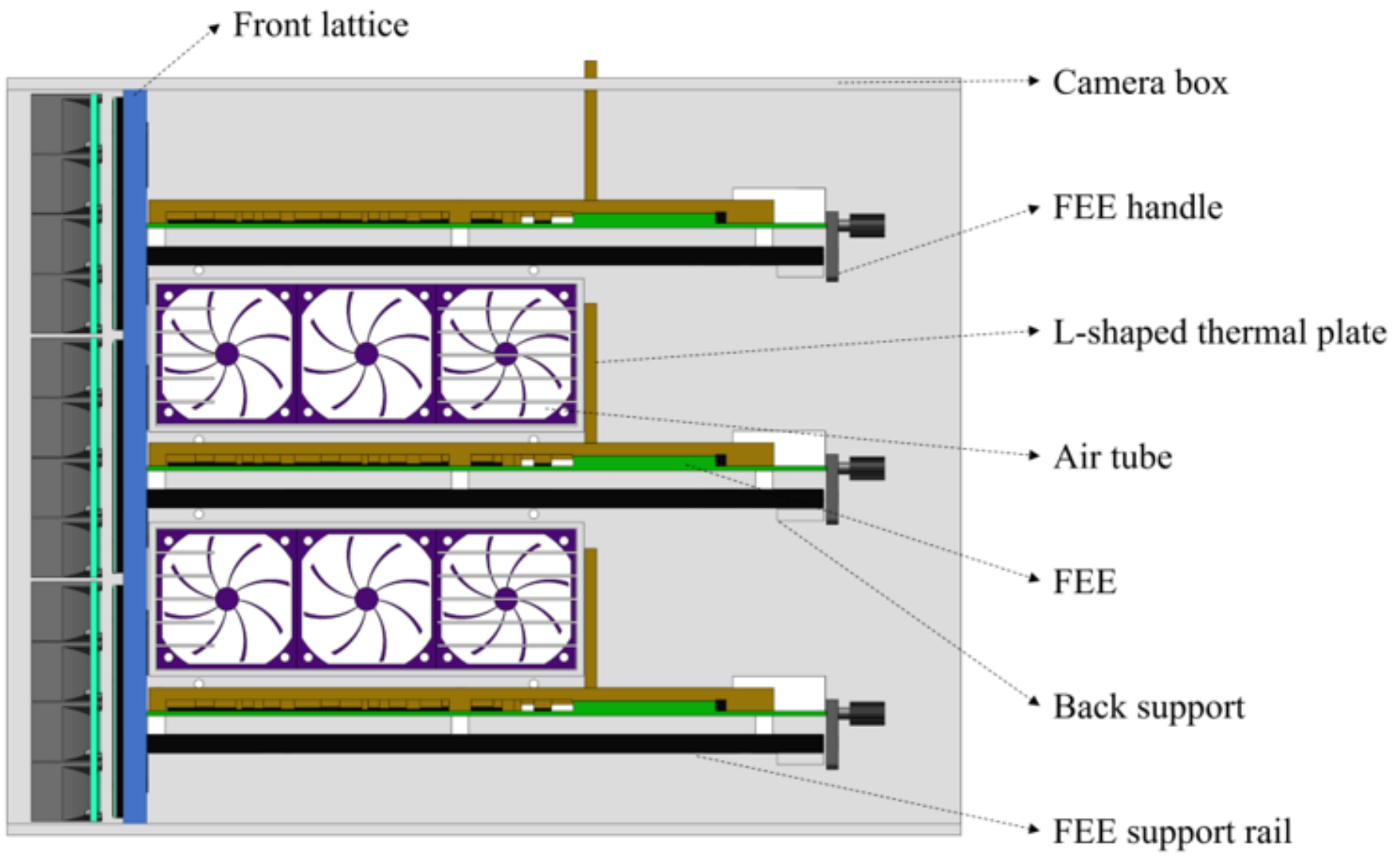

**Figure 2.** Schematic of the cooling structure of the LACT camera [5].

The LACT camera contains 101 sub-clusters, and the compact pixel arrangement limits the space available for each FEE. The design integrates 16 channels of SiPM operating voltage adjustment and high-speed waveform sampling within the available board dimensions. In addition, the low air density at high altitude limits convective heat transfer. Reducing electronics power consumption helps ease the cooling load and limit the temperature rise inside the camera. The FEE is also designed to operate over a wide temperature range while maintaining stable gain compensation and high-speed waveform sampling. These space, power consumption and temperature constraints form the basis of the FEE design. Its overall architecture and circuit implementation are described below.

## 2. Circuit design

### 2.1 Gain compensation circuit

The compensation circuit used in the FEE to compensate for the SiPM gain temperature drift follows our previously developed design. Its detailed implementation, calibration method and performance tests are reported in [6]. As shown in figure 3, the compensation circuit comprises temperature monitoring, operating voltage adjustment and operating voltage monitoring.

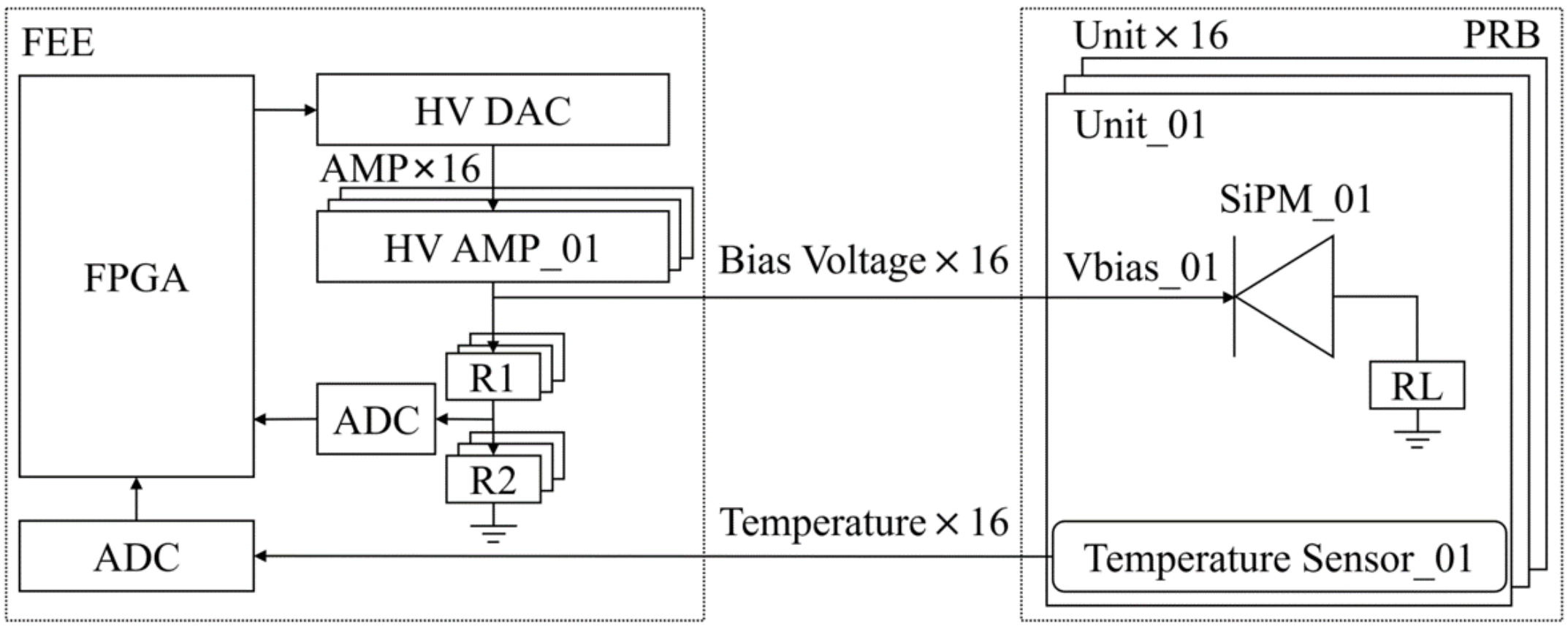

**Figure 3.** Block diagram of the circuit for compensating the SiPM gain temperature drift.

The 16 analog temperature sensors on the PRB measure the operating temperatures of the corresponding SiPMs. Each sensor is located behind its SiPM, and its analog output is digitized by a 16-channel, 16-bit AD4695 ADC on the FEE. To stabilize the SiPM gain, the voltage-temperature compensation relation is determined from the previously calibrated dependence of gain on temperature and operating voltage. The slow control system uses this relation and the measured temperature of each channel to determine the operating voltage setting for each SiPM.

The voltage regulation circuit consists of a 16-channel, 16-bit DAC81416 high-voltage digital-to-analog converter and 16 LTC6090 high-voltage operational amplifiers. The slow control system configures the DAC81416 through a field-programmable gate array (FPGA) according to the operating voltage setting of each channel. Each DAC81416 output is amplified by a factor of 2 by an LTC6090 to provide the operating voltage for the corresponding SiPM. The DAC81416 is configured for an output range of 0 V to 40 V. After amplification by a factor of 2, the voltage adjustment step is approximately 1.22 mV. LACT currently uses the customized HAMAMATSU-S17351 SiPM, whose recommended operating voltage range is 57.5 V to 63.5 V. The LTC6090 operates from a single 65 V supply. The SiPM operating voltage adjustment range is 1 V to 64 V, leaving a margin of approximately 1 V above the lower supply rail and below the upper supply rail.

To monitor the actual SiPM operating voltage, each channel uses a divider formed by R1 (150 kΩ) and R2 (10 kΩ) to reduce the voltage by a factor of 16. The divided voltage is digitized by a second 16-channel, 16-bit AD4695. With an ADC reference voltage of 4.096 V, the voltage measurement step referred to the SiPM terminals is 1 mV. The slow control system compares the voltage readback with the set value to check the operating status of the regulation circuit and the SiPM.

### 2.2 High-speed waveform sampling circuit

The LACT camera is designed to cover Cherenkov light signals from 5 photoelectrons (pe) to 8000 pe, corresponding to a ratio of 1600 between the maximum and minimum photoelectron counts [4]. The previous study of shaping circuits specified a pulse integral area resolution below 5% at 1000 pe [4]. To meet this requirement, the OPA892 gain is set to 25.26. With this gain, the pulse amplitude at the OPA892 output is 3100 mV for a signal of 8000 pe. Figure 4 shows the pulse waveform from a HAMAMATSU-S17351 SiPM after OPA892 amplification at a light intensity of approximately 1000 pe. The rise time is approximately 8.90 ns (measured from 5% to 95%), and the full width at half maximum (FWHM) is approximately 35.62 ns.

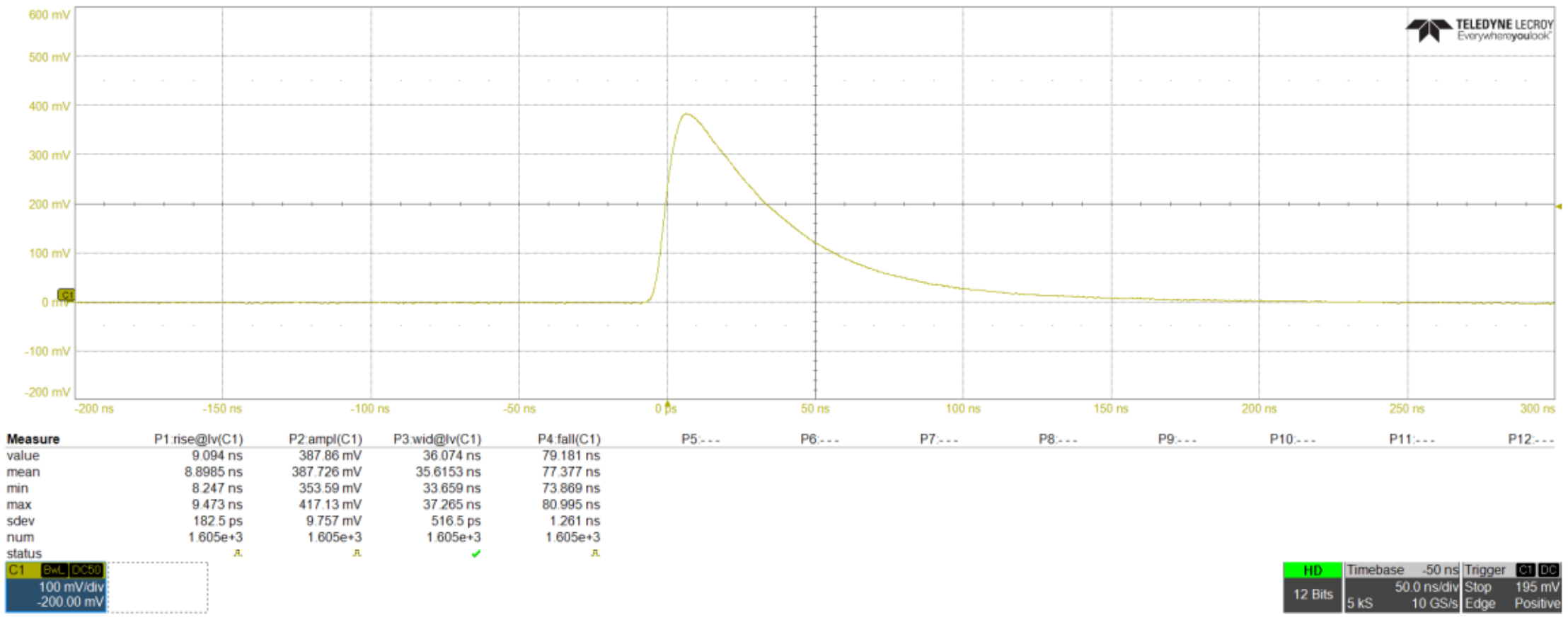

**Figure 4.** SiPM pulse waveform after OPA892 amplification at a light intensity of approximately 1000 pe.

The high-speed waveform sampling circuit is designed for pulse amplitudes from 0 mV to 3100 mV, referenced to the OPA892 output. To reduce the contribution of electronics noise to the light intensity measurement, LACT requires an integral area resolution below 1% for electrical input pulses equivalent to a 1000 pe signal. At the OPA892 output, the equivalent electrical pulse is chosen to have a rise time and FWHM close to those of the waveform in figure 4. Its amplitude is set to one eighth of the full-scale pulse amplitude, namely 387.5 mV.

There are two main approaches to high-speed waveform sampling: analog sampling and storage using a switched capacitor array (SCA), and direct digitization using a fast analog-to-digital converter (FADC). The DRS4 is a representative SCA sampling chip [8] and has been used in the readout electronics of the MAGIC telescopes and the Large-Sized Telescope of the Cherenkov Telescope Array [9, 10]. The DRS4 sequentially switches sampling capacitors to store the input voltage at different times. After a trigger arrives, the stored analog waveform is digitized sequentially by an external ADC. The DRS4 provides waveform sampling at rates of billions of samples per second with relatively low chip power consumption. However, the external ADC increases the component count and power consumption. In this readout mode, sampling in the corresponding channel is interrupted during waveform readout and digitization, introducing readout dead time. In addition, the approximately 1 V input range of the DRS4 requires substantial attenuation of a 3100 mV pulse. This attenuation makes it more difficult to maintain adequate resolution for small signals.

The FADC approach directly converts the input waveform into digital data. It avoids the post-trigger digitization step required in the SCA approach and is suitable for continuous waveform sampling. However, high-speed ADCs have relatively high power consumption. For example, the AD9643 from Analog Devices is a dual-channel, 14-bit ADC with a power consumption of 392.5 mW per channel at a sampling rate of 250 MSPS [11]. For the large number of channels and limited installation space in the LACT readout electronics, the combined power consumption of the sampling devices creates a significant cooling load. It is necessary to reduce per-channel power consumption and increase the level of integration while meeting the waveform sampling requirements.

Considering both the sampling density across the pulse waveform and the electronics power consumption, a sampling rate of 250 MSPS is adopted. This corresponds to a sampling interval of 4 ns, with approximately nine sampling intervals across the pulse FWHM, providing a basis for measuring the pulse integral area. The AC9628, developed by HiSilicon, is selected for the high-speed waveform sampling circuit [12]. Each device integrates 16 ADCs, each with a maximum sampling rate of 125 MSPS. The design uses two AC9628 devices, each configured for eight analog inputs and 14-bit operation. Within each device, two ADCs operating at 125 MSPS sample the same analog input in an interleaved manner, providing an effective sampling rate of 250 MSPS. The two devices together sample the pulse signals from all 16 SiPM channels. In the 14-bit, 125 MSPS configuration with JESD204B output, the power consumption of one AC9628 is 576 mW, corresponding to 72 mW per 250 MSPS input channel. The low power consumption per channel and the multichannel integration help implement a compact, low-power sampling circuit. In addition, the AC9628 has a differential full-scale input range of 2 V. Compared with the approximately 1 V input range of the DRS4, this reduces the attenuation requirement and provides more margin in matching the signal ranges.

The high-speed waveform sampling circuit is shown in figure 5. After shaping on the PRB, the SiPM pulse signals are sent to the FEE and amplified by the OPA892, with a maximum pulse amplitude of approximately 3100 mV. A THS4541 fully differential amplifier converts the single-ended OPA892 output into a differential signal (DS), with a differential gain of 0.613. A DC offset of 1550 mV is applied to the other THS4541 input. This produces an ADC differential input voltage range of approximately −950 mV to +950 mV, matching the AC9628 differential input range of −1 V to +1 V.

The digitized waveform data are transmitted from the AC9628 devices to the FPGA through JESD204B interfaces. Each AC9628 uses four high-speed serial differential links with current-mode logic (CML). The FPGA on the FEE is an XC7K325TFFG900-2, which provides 16 GTX transceiver channels. Eight of these channels receive the digitized waveforms from the 16 analog channels of the two AC9628 devices.

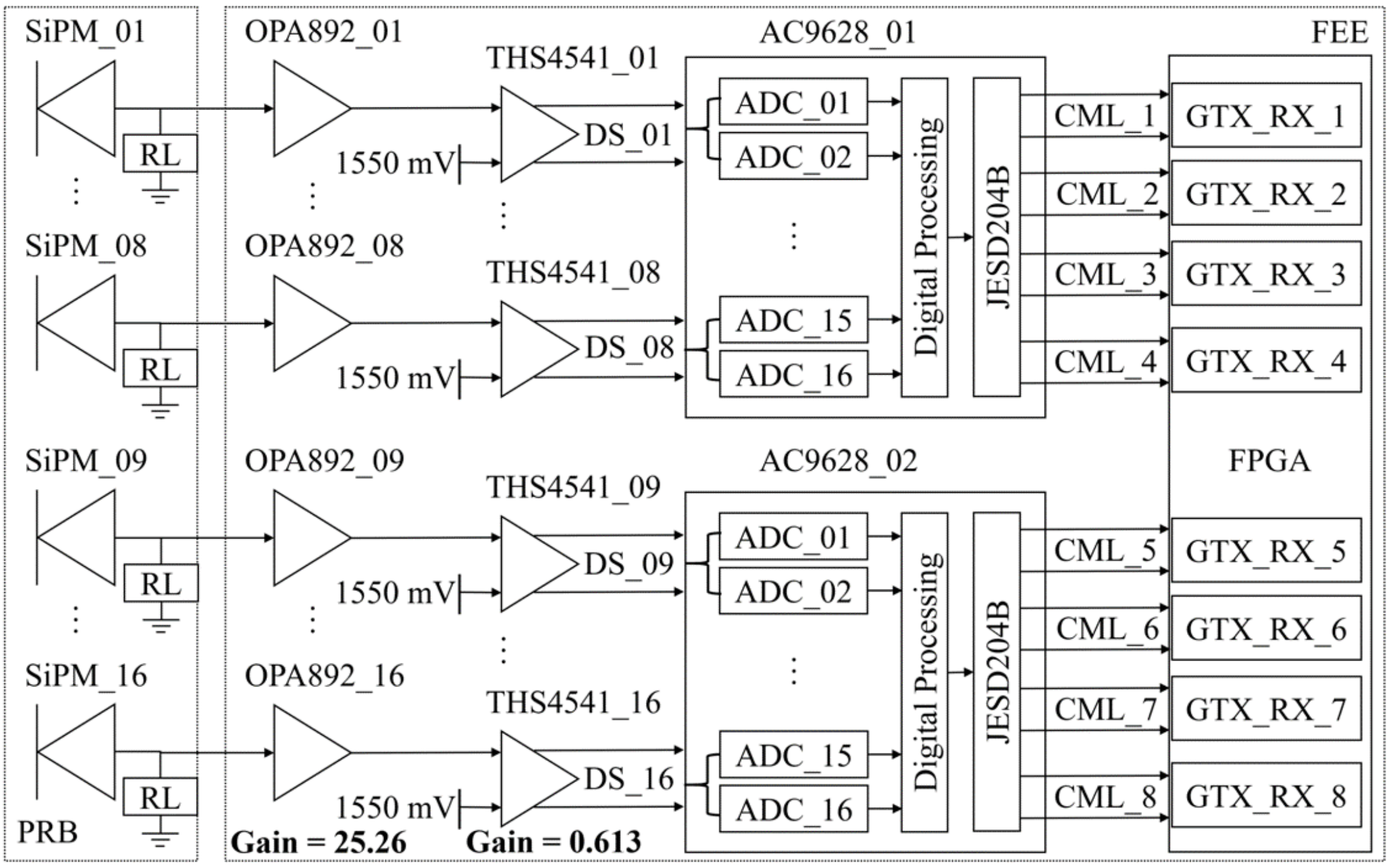

**Figure 5.** Block diagram of the high-speed waveform sampling circuit.

## 2.3 Board layout

The layout of the main components on the FEE is shown in figure 6. The SiPM pulse signals enter the FEE through the PRB connector and pass sequentially through OPA892 amplification, THS4541 differential conversion and AC9628 digitization. These devices are placed close together in signal flow order to shorten the analog signal paths. An ADP125 provides a DC offset voltage of 1550 mV for the THS4541 amplifiers. An LMK04828 and its associated CVHD-950X-125 provide high-precision clock and system reference signals to the AC9628 devices and the FPGA for waveform sampling and high-speed digital data transmission.

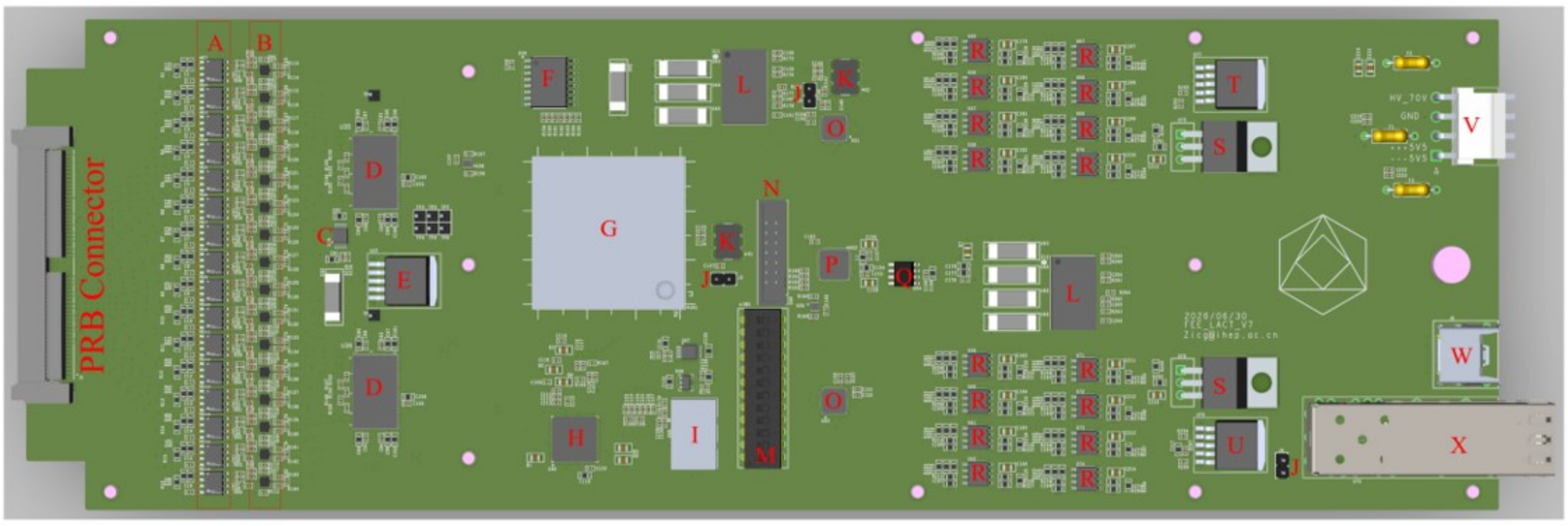

| | | | | | |
|---|---|---|---|---|---|
| A: OPA892 | B: THS4541 | C: ADP125 | D: AC9628 | E: LT1764-1.8 | F: Configuration Flash |
| G: FPGA | H: LMK04828 | I: CVHD-950X-125 | J: Jumper | K: SiT9121AI-2D1-33E125 | L: LTM4644 |
| M: DIP Switch | N: JTAG Connector | O: AD4695 | P: DAC81416 | Q: ADR4525 | R: LTC6090 |
| S: TL783 | T: LT1764 | U: LT1185 | V: Power Connector | W: Data Connector | X: SFP |

**Figure 6.** Layout of the FEE.

The SiPM operating voltage regulation circuit mainly consists of the DAC81416, an ADR4525 voltage reference and 16 LTC6090 amplifiers. The relatively high power consumption of the DAC81416 and LTC6090 amplifiers leads to localized heating in this section. The high-voltage devices are therefore placed near the vertical section of the L-shaped thermal plate to facilitate heat conduction to the air tube wall. Two AD4695 devices digitize the divided operating voltages and the analog outputs of the temperature sensors on the PRB, respectively. They provide monitoring of the actual SiPM operating voltages and temperatures.

External power enters through the power connector. Two LTM4644 devices and linear regulators supply the different circuit blocks. The data connector provides communication between the FEE and the trigger and data acquisition board, which performs trigger decisions and data processing for one camera. The FEE also includes a JTAG connector for FPGA programming and a flash memory for storing the FPGA configuration.

To support standalone testing and calibration before installation in the camera, the FEE retains two local clock sources and a small form-factor pluggable (SFP) communication interface. Two SiT9121 clock sources provide the local system clock and the GTX reference clock required for SFP communication, respectively. The FEE communicates with a host computer through the SFP interface for standalone test control and data transfer. A DIP switch sets the board identifier used to identify each FEE during testing and calibration. This identifier is also used to associate the calibration files with the corresponding board.

After the FEE is installed in the camera, the trigger and data acquisition board supplies the system clock. The FEE exchanges control and waveform data with the trigger and data acquisition board through the data connector. The two local clock sources and the SFP communication link are not needed. Jumpers are provided in the power supply paths of the two clock sources and the SFP module. Opening these jumpers when the FEE is installed in the camera further reduces its operating power consumption.

## 3. Performance tests

### 3.1 Gain compensation circuit performance

The SiPM gain compensation circuit used in the FEE was characterized in our previous study. The test methods and detailed results are reported in [6] and are briefly summarized here. After calibration, the integral nonlinearities of the voltage regulation and monitoring circuits are 0.0012% and 0.0011%, respectively. At a regulation circuit output voltage of 64 V, the root mean square (RMS) noise is 2.59 mV, measured with a bandwidth of 20 MHz. Over a temperature range of −25 °C to 45 °C, the cumulative output voltage drift is 12.13 mV. The maximum absolute monitoring error over the same range is 34.68 mV. A linear correction based on the temperature dependence of the monitoring error reduces it to 0.98 mV. The previous study tested gain compensation for three SiPM models: S14466, S14520 and J40035. For the three SiPMs, the compensated gain drift remains below 2% from −20 °C to 30 °C, demonstrating the applicability of the scheme to different SiPMs.

### 3.2 High-speed waveform sampling circuit performance

#### 3.2.1 DC response and baseline noise

The test setup for the high-speed waveform sampling circuit is shown in figure 7. Small-signal tests require differential pulse amplitudes as low as 1 mV at the ADC input. The combined voltage gain of the OPA892 and THS4541 is approximately 15.48, so the corresponding pulse amplitude at the FEE input is only approximately 64.6 μV. The AFG31252 function generator has an amplitude setting step of 1 mV. An R&S step attenuator from Rohde & Schwarz is therefore used to attenuate larger signals to the required levels before they are delivered to the FEE through the adapter board.

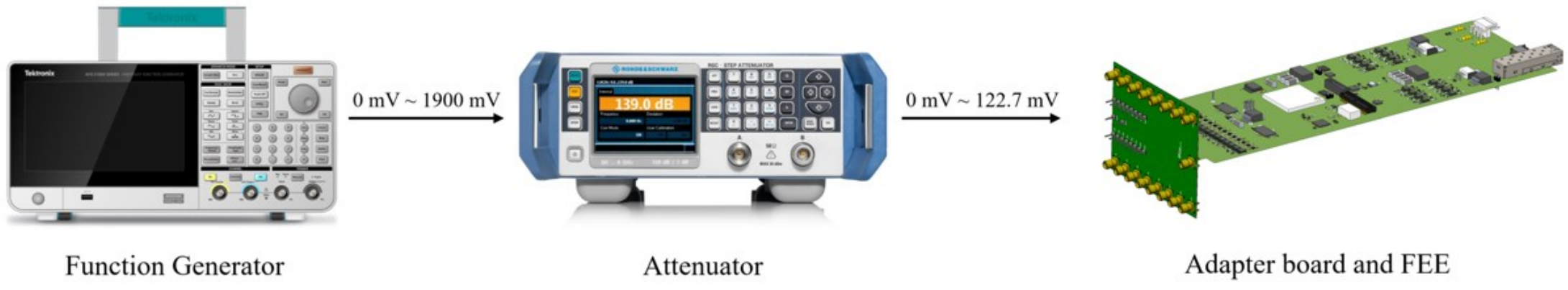

**Figure 7.** Test setup for the high-speed waveform sampling circuit.

The DC or pulse signal from the function generator is attenuated and delivered to the FEE through the SMA connector on the adapter board. It then passes through OPA892 amplification, THS4541 differential conversion and AC9628 digitization. The FPGA reads out the sampled data and transmits them to the host computer through the SFP interface for analysis. This setup is used to measure the DC response linearity, baseline noise, interchannel crosstalk and pulse integral area resolution.

To characterize the DC response, the DC output voltage of the function generator is increased from 0 mV to 1900 mV in steps of 100 mV. After attenuation, amplification and differential conversion, the corresponding differential voltage at the AC9628 input ranges from approximately −950 mV to +950 mV. At each DC voltage setting, 10000 waveforms are recorded, each containing 256 samples, giving a total of 2.56 million samples. The mean of all samples is calculated at each voltage setting. The mean sampled value as a function of the generator DC output voltage and its linear fit are shown in figure 8(a). At each test point, the difference between the measured data and fitted values is divided by the fitted output span to obtain the normalized deviation, as shown in figure 8(b). The maximum absolute deviation occurs at a generator output of 0 mV, giving an integral nonlinearity (INL) of 0.0709%.

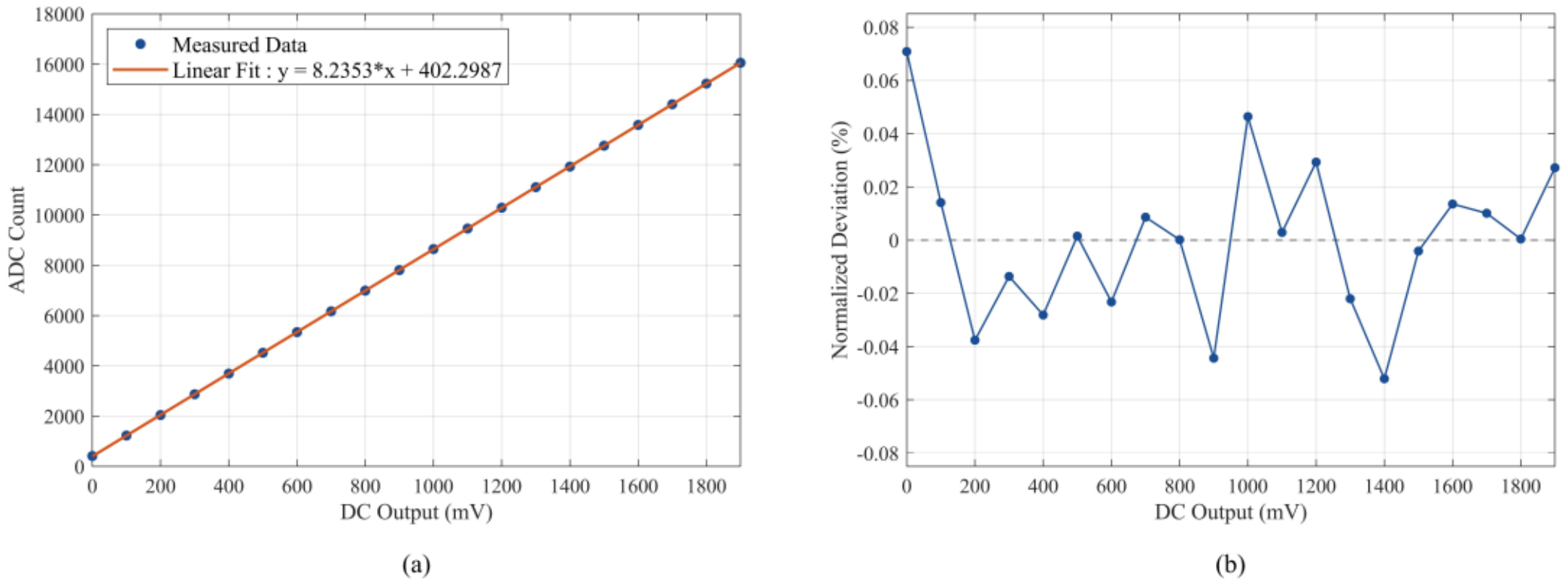

**Figure 8.** DC response of the high-speed waveform sampling circuit.

The baseline noise is measured with the function generator set to a DC output of 0 mV. Again, 10000 waveforms of 256 samples each are recorded, giving a total of 2.56 million samples. The AC9628 achieves 250 MSPS sampling by interleaving two sub-ADCs operating at 125 MSPS. Odd samples and even samples originate from different sub-ADCs. Because the two sub-ADCs may have different baselines, the mean of the odd sample sequence and the mean of the even sample sequence are subtracted separately when calculating the noise. The RMS of the remaining fluctuations characterizes the baseline noise. Figure 9(a) shows the baseline noise distribution for one channel, with an RMS of 5.7093 ADC counts.

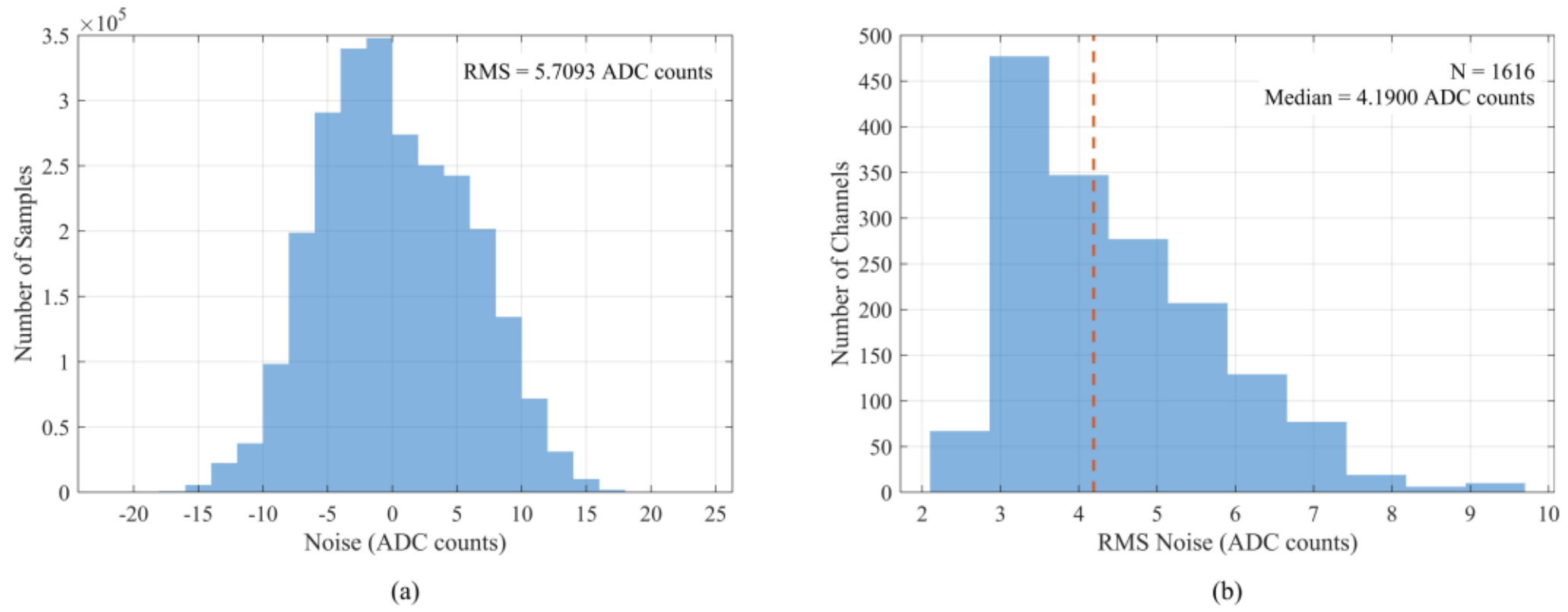

**Figure 9.** Baseline noise distributions under zero-input conditions.

To evaluate the effect of adjacent-channel crosstalk on the baseline noise, the same measured channel is kept at zero input while a 25 MHz sinusoidal signal is applied to the ADC input of an adjacent channel. The differential voltage varies from −950 mV to +950 mV, corresponding to a peak-to-peak amplitude of 1900 mV, close to the ADC full scale. With the same data processing method, the RMS noise of the measured channel is 6.2791 ADC counts. This is an increase of 0.5698 ADC counts relative to the result without the adjacent-channel excitation.

The same zero-input baseline noise test is performed on all 1616 channels of the 101 FEE boards used in one camera. The distribution of the channel RMS values is shown in figure 9(b). Its median is 4.1900 ADC counts, representing the typical FEE baseline noise level.

### 3.2.2 Pulse response

To characterize the pulse response and integral area resolution of the high-speed waveform sampling circuit, a function generator produces equivalent electrical pulses that simulate SiPM signals. The pulse amplitude is set to 237 mV and the pulse repetition frequency is set to 1 kHz. The generator output waveform measured with an oscilloscope is shown in figure 10. Its amplitude is 237.43 mV, its 5% to 95% rise time is 8.97 ns, and its FWHM is 37.42 ns. The rise time and FWHM are close to those of the SiPM waveform in figure 4. After attenuation, amplification and differential conversion in the test signal path, the equivalent pulse produces a differential pulse amplitude of approximately 237 mV at the AC9628 input. Referred to the OPA892 output, this amplitude corresponds to approximately 387 mV, equivalent to a signal level of approximately 1000 pe.

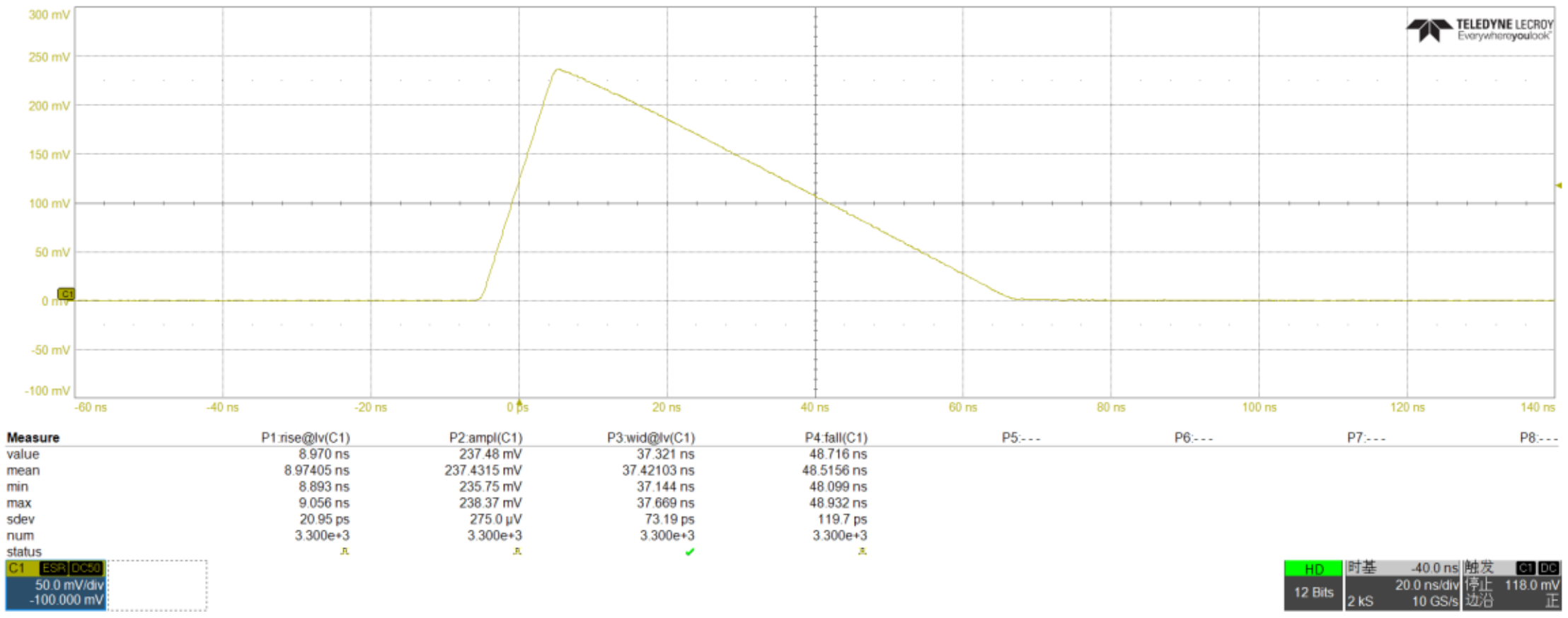

**Figure 10.** Equivalent electrical pulse waveform at the function generator output.

The channel shown in figure 9(a) is used for the pulse test. The FEE samples 10000 equivalent pulses at 250 MSPS. An example of a raw sampled pulse waveform is shown in figure 11(a). For each waveform, the first 50 samples are selected as the baseline interval. The means of the odd samples and even samples in this interval are calculated separately to estimate the baselines of the two sub-ADCs. These baselines are then subtracted from the corresponding odd samples and even samples of the full waveform. The corrected values at sample points 50 through 100 are summed, and this discrete sum is used to represent the pulse integral area. The integral area distribution is shown in figure 11(b). Its mean is 17131.7582 ADC counts, and the RMS fluctuation of the integral area is 46.6834 ADC counts. The integral area resolution is defined as the ratio of this RMS to the mean integral area. The resolution for the equivalent electrical pulses is 0.2725%, meeting the design requirement of less than 1%.

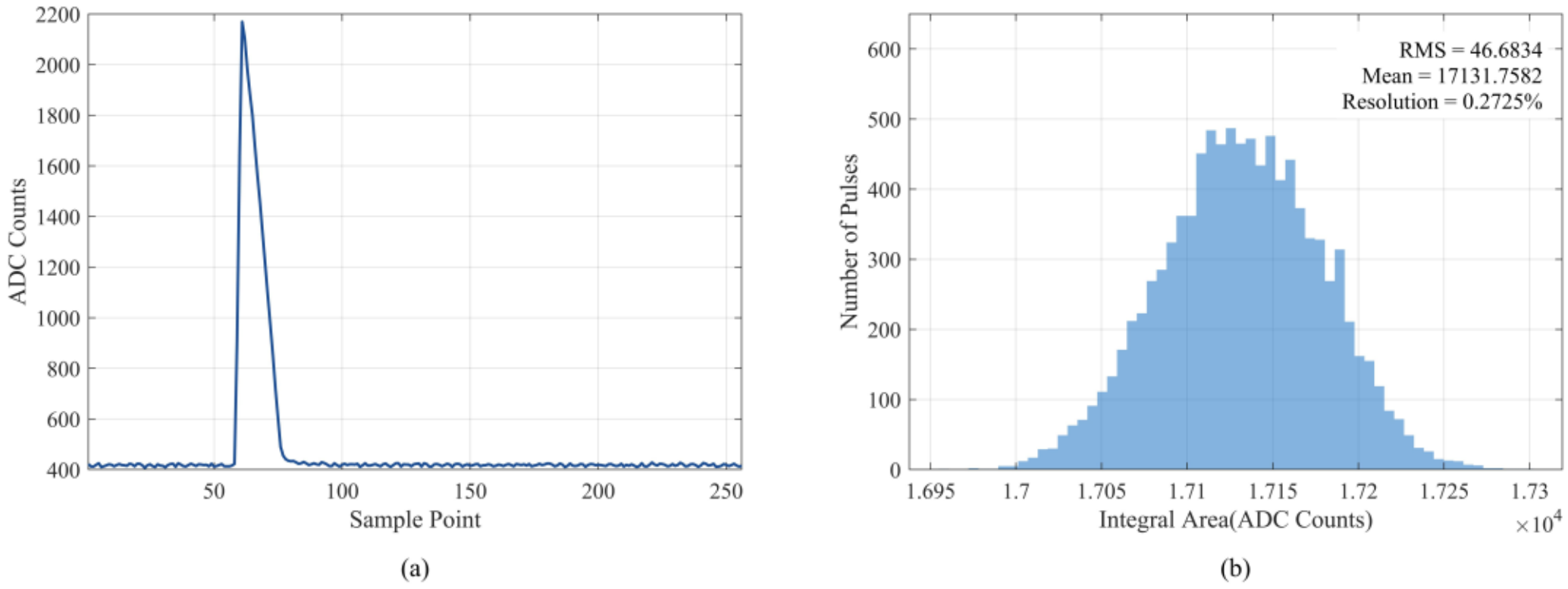

**Figure 11.** Sampled waveform and integral area distribution for equivalent electrical pulses.

To characterize the response at different pulse amplitudes, the generator pulse amplitude is increased from 1 mV to 1900 mV while the pulse shape and pulse repetition frequency are kept unchanged. At each amplitude, 10000 pulses are sampled. The same baseline subtraction method and integration window are used to calculate the mean integral area and its resolution. The mean integral area as a function of generator pulse amplitude and its linear fit are shown in figure 12(a). Using the same normalization method as in the DC response test, the INL of the integral area is 0.04820% over the tested amplitude range. Figure 12(b) shows the integral area resolution as a function of pulse amplitude. The resolution is below 1% at every tested amplitude from 100 mV to 1900 mV.

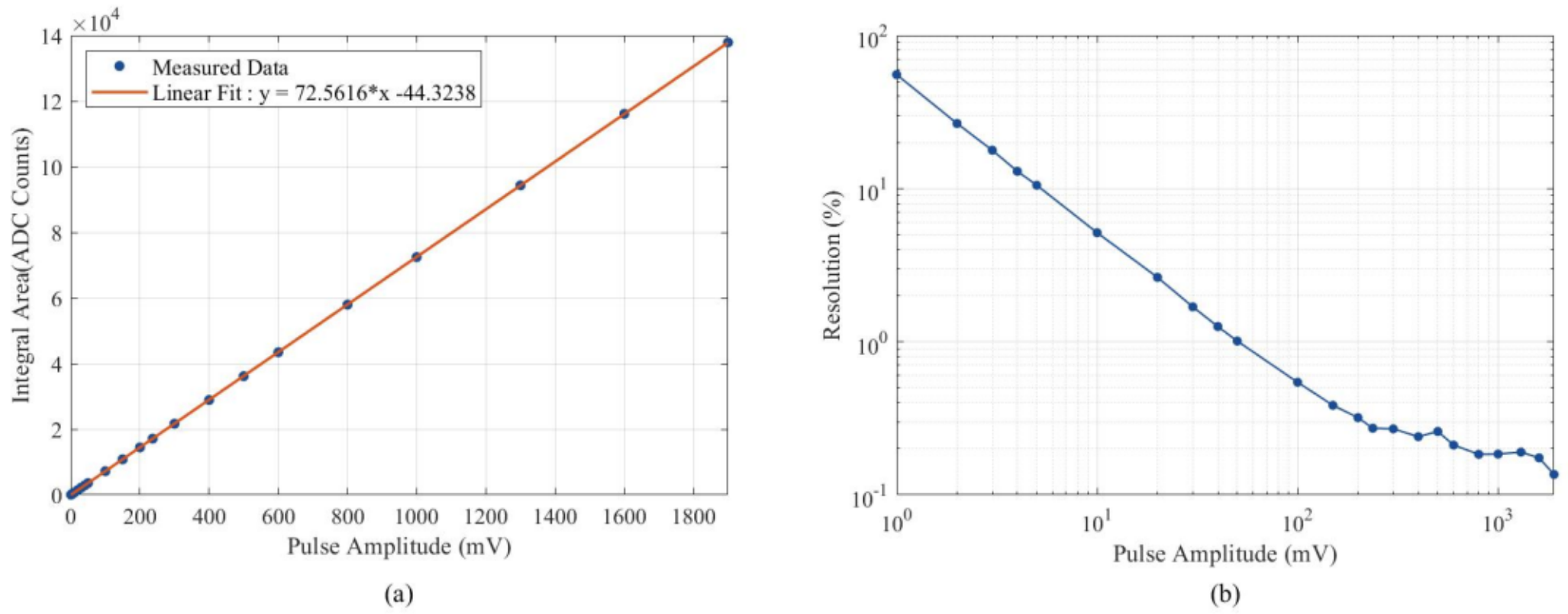

**Figure 12.** Integral area response and resolution for equivalent electrical pulses of different amplitudes.

To test the FEE response to actual SiPM signals, a PRB carrying HAMAMATSU-S17351 SiPMs is assembled with the FEE as shown in figure 1. At a light intensity of approximately 1000 pe, the FEE samples 10000 pulses from the PRB. An example of a raw sampled waveform is shown in figure 13(a). The same correction method is used to subtract the baselines of the odd sample sequences and the even sample sequences separately for each waveform. Because the SiPM pulses have a longer decay tail, the integration window is extended to sample points 50 through 150. The corrected samples in this window are summed to obtain the integral area. The integral area distribution is shown in figure 13(b). At approximately 1000 pe, the integral area resolution measured in the combined PRB and FEE test is 3.5441%.

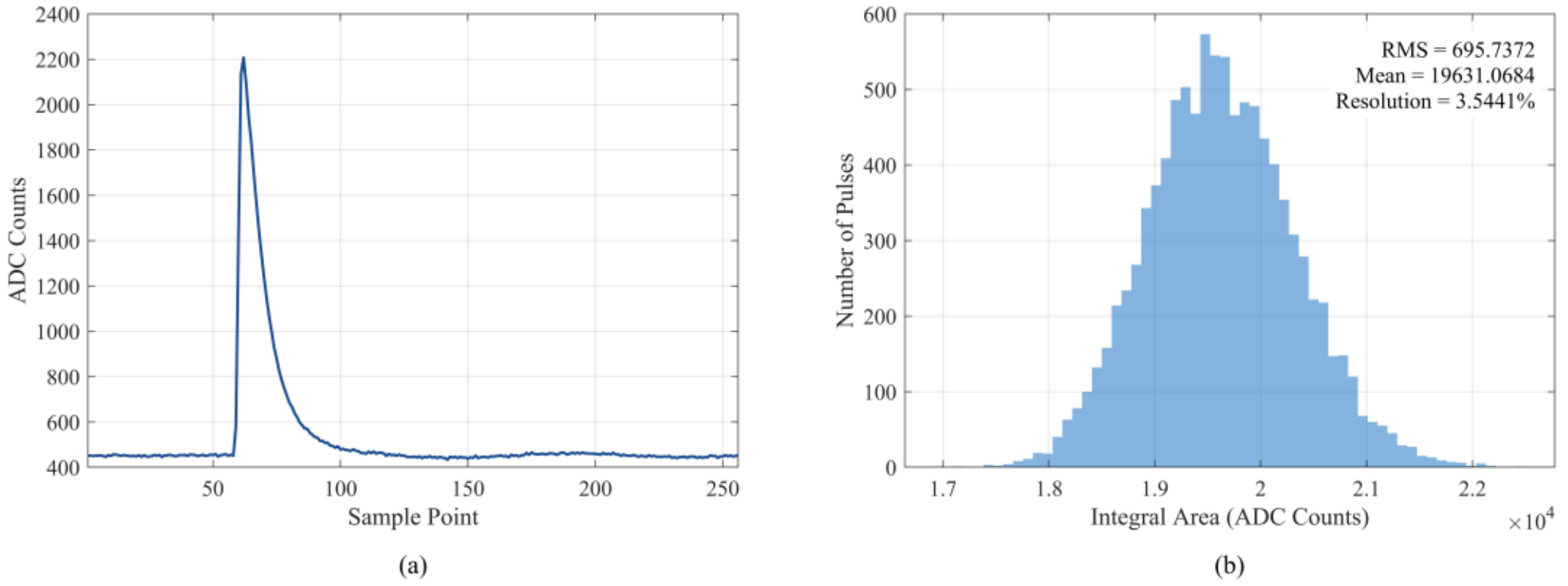

**Figure 13.** Sampled SiPM waveform and area distribution at a light intensity of approximately 1000 pe.

### 3.2.3 Temperature dependence

The temperature dependence of the high-speed waveform sampling circuit is measured using the connections in figure 7, with the test adapter board and FEE placed in a temperature chamber. The function generator produces equivalent electrical pulses with an amplitude of 237 mV. After attenuation, the test signal is delivered to the chamber through a long cable. The chamber temperature is varied from −25 °C to 45 °C. At each temperature point, 10000 equivalent pulses are sampled after the temperature sensor reading inside the chamber has stabilized at the set value. The mean integral area and its resolution at each temperature are calculated using the same correction method and integration window as those used for figure 11.

The temperature dependence of the mean integral area is shown in figure 14(a). The right vertical axis gives the integral area normalized to the result at 20 °C. The mean integral area increases with temperature, and the normalized values range from 0.9923 to 1.0063. Once the temperature response of the sampling circuit is incorporated into the overall system temperature calibration, the effect of temperature drift on the overall response can be compensated by adjusting the SiPM operating voltage. The integral area resolution as a function of temperature is shown in figure 14(b). It ranges from 0.3477% to 0.6890% and remains below 1% at all tested temperatures.

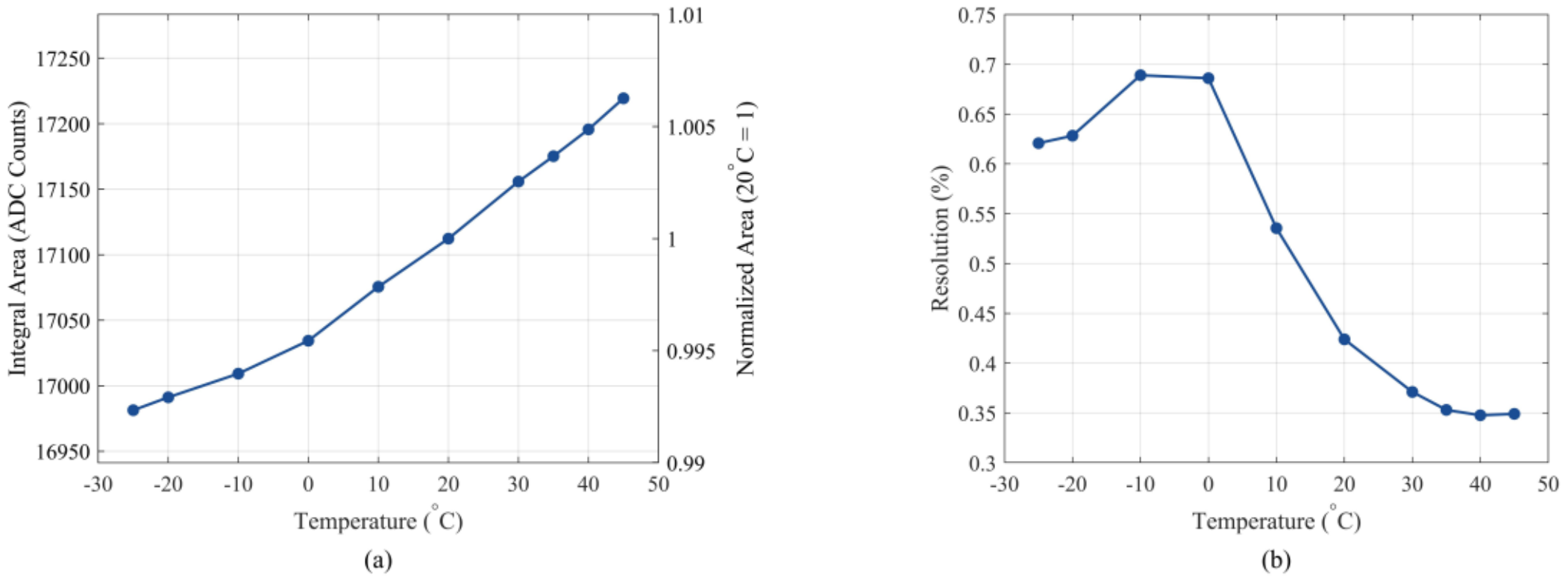

**Figure 14.** Temperature dependence of the high-speed waveform sampling circuit.

## 4. Conclusions and discussion

A front-end electronics board has been developed to meet the requirements for SiPM gain stabilization and pulse signal measurement in the LACT camera. A single board integrates 16 channels of independent SiPM operating voltage adjustment, operating temperature and operating voltage monitoring, and 14-bit waveform sampling at 250 MSPS. The FEE uses a previously validated SiPM gain compensation scheme, which limits the gain drift of the tested SiPMs to below 2% over the temperature range from −20 °C to 30 °C.

The high-speed waveform sampling circuit has an INL of 0.0709% for the mean sampled value as a function of the generator DC output voltage. The INL of the integral area as a function of pulse amplitude is 0.04820%. Baseline noise measurements on the 101 FEE boards for one camera give a median RMS of 4.1900 ADC counts across 1616 channels. The integral area resolution is 0.2725% for electrical pulses equivalent to a 1000 pe signal and is 3.5441% for actual SiPM pulse signals at a light level of approximately 1000 pe. Temperature tests with equivalent electrical pulses show that, over the temperature range from −20 °C to 30 °C, the mean integral area changes by −0.71% to +0.25% relative to the result at 20 °C. The integral area resolution is below 1% at all tested temperatures.

During standalone testing, the total power consumption of one FEE is 18.86 W, comprising 7.09 W in the high-voltage section and 11.77 W in the low-voltage section. For operation in the camera, the power supplies to the two local clock sources and the SFP module are disconnected. The measured power consumption of the low-voltage section decreases by 1.05 W, reducing the total to 17.81 W. This design retains the convenience of standalone testing and calibration while meeting the need for low power consumption during camera operation. The developed FEE provides a front-end electronics solution for multichannel SiPM gain stabilization and pulse signal readout in the LACT camera.

## Acknowledgments

The authors would like to express gratitude to all members of the LACT group. This work is supported in part by the National Natural Science Foundation of China under grant No. 12342502.